\documentclass[10pt,twocolumn,english,aps,pra,showpacs]{revtex4}
\usepackage{amsfonts}
\usepackage[T1]{fontenc}
\usepackage[latin9]{inputenc}
\usepackage{amsmath}
\usepackage{graphicx}
\usepackage{amssymb}
\usepackage{esint}
\usepackage{multirow}
\usepackage{booktabs}
\usepackage{tabu}

\makeatletter
\@ifundefined{textcolor}{}
{%
 \definecolor{BLACK}{gray}{0}
 \definecolor{WHITE}{gray}{1}
 \definecolor{RED}{rgb}{1,0,0}
 \definecolor{GREEN}{rgb}{0,1,0}
 \definecolor{BLUE}{rgb}{0,0,1}
 \definecolor{CYAN}{cmyk}{1,0,0,0}
 \definecolor{MAGENTA}{cmyk}{0,1,0,0}
 \definecolor{YELLOW}{cmyk}{0,0,1,0}
 }

\begin{document}

\title{Direct state reconstruction with coupling-deformed pointer observables}

\author{Xuanmin Zhu{$^{1}$}}\email{zhuxuanmin2006@163.com}
\author{Yu-Xiang Zhang$^{2,3,4}$}\email{iyxz@mail.ustc.edu.cn}
\author{Shengjun Wu$^{4}$}\email{sjwu@nju.edu.cn}

 \affiliation{$^{1}$ School of Physics and Optoelectronic Engineering, Xidian University, Xi'an 710071, China \\
 $^{2}$Hefei National Laboratory for Physical Sciences at Microscale, University of Science and Technology of China, Hefei, Anhui 230026, China\\
 $^{3}$ The CAS Center for Excellence in QIQP and the Synergetic Innovation Center
for QIQP, University of Science and Technology of China, Hefei, Anhui 230026, China \\
  $^{4}$Kuang Yaming Honors School, Nanjing University, Nanjing, Jiangsu 210093, China}

\date{\today}
\begin{abstract}
    Direct state tomography (DST) using weak measurements has received wide attention.
    Based on the concept of coupling-deformed pointer observables presented by Zhang \emph{et al}.[Phys. Rev. A \textbf{93}, 032128 (2016)],
    a modified direct state tomography (MDST) is proposed, examined, and compared with other typical state tomography schemes. MDST has exact validity for measurements of any strength.
    We identify the strength needed to attain the highest efficiency level of MDST by using statistical theory.
    MDST is much more efficient than DST in the sense that far fewer samples are needed
    to reach DST's level of reconstruction accuracy. Moreover, MDST has no inherent bias when compared to DST.
\end{abstract}

\maketitle

\section{Introduction}

Though the quantum no-clone theorem prohibits the perfect estimation of the unknown state of
a single quantum system~\cite{ncl1,ncl2}, state reconstruction is possible via repeatedly measuring
an ensemble of identical systems, a process usually called quantum state tomography (QST).
Besides the standard QST strategies~\cite{qst1,qst2,qst3,qst4,qst5,qst6,qst7,qst8}, a novel tomography strategy, conventionally called \emph{direct state tomography} (DST), or weak-value tomography, has been widely investigated both theoretically and experimentally ~\cite{dsm1,dsm2,dsm3,dsm4,dsm5,dsm7,dimension,comwave,high}.

DST is based on the quantum weak value theory introduced by Aharonov, Albert and Vaidman (AAV) in 1988~\cite{aav}.
In DST, each element of the unknown density operator is proportional to a single \emph{weak value},
if we choose the appropriate variables in weak measurements \cite{dsm3}.
In this way, the wave function at each point can be determined directly, without the global
inversion required in standard QST \cite{dsm7}.
Because of this feature and the simplicity of experimental implementation,
DST can be conveniently realized and is probably the only choice for
the tomography of high-dimensional states \cite{dimension,comwave,high}.

However, as pointed out in~\cite{mac}, DST suffers from the disadvantages,
including low efficiency and systematical reconstruction bias.
It is conceivable that these flaws mainly stem from the AAV's weak-value formalism that built on
first-order perturbation of the measurement strength~\cite{wk1,wk2,wk3,wk4}.
That is, very weak coupling strength causes the low efficiency of DST~\cite{mac};
approximations produce unavoidable bias~\cite{mac}.

Recently, we have constructed a new framework for quantum measurements with postselection~\cite{zhan}.
In our formalism, weak value information can be generated exactly with measurements of any strength,
provided that two \emph{coupling-deformed(CD) pointer observables} are read on the pointer of the quantum measuring device~\cite{zhan}.
The general formula for determining the CD observables was also given in \cite{zhan}.
In particularly, when a single qubit is used as the measuring device,
it is also reported in Refs.~\cite{vall,zou} that weak values can be obtained from stronger
measurements by standard state tomography of the qubit device.

In this paper, we propose a modified DST that works by applying
the CD observables \cite{zhan} in state tomography; it will be denoted as MDST hereafter.
Since MDST is valid over the full range of measurement strength,
we will determine the optimal measurement strength at which MDST attains its highest efficiency.
Then, through Monte Carlo simulations,
we demonstrate that the efficiency of MDST is much higher than that of DST.
That is, to reach the same level of reconstruction accuracy, MDST needs far fewer samples.
Furthermore, MDST is also compared with SU(2) tomography,
one of the most efficient stand QST~\cite{qst1}.

This paper is organized as follows.
We briefly review SU(2) tomography and DST in Sec. \ref{sec2}.
In Sec. \ref{sec3}, MDST is studied and the optimal coupling strength
is given.  In section \ref{sec4}, the performance of
different state reconstruction strategies are compared using Monte Carlo simulation results.
A short conclusion is presented in Sec. \ref{sec5}.

\section{Strategies for quantum state reconstruction}\label{sec2}

In this section, we briefly review SU(2) tomography, and the original
direct state tomography (DST).

\subsection{SU(2) tomography}
SU(2) tomography is a well-established state reconstruction technique.
It is based on the formula~\cite{qst1}
\begin{equation}\label{e2}
\rho_{in}=\int dg R^{\dagger}\left(g \right) \mathrm{tr}[\rho_{in} R(g)],
\end{equation}
where $R\left(g\right)$ is the unitary irreducible square-integrable representation of
a tomographic group $G$, $g\in G$.
The derivation of Eq. (\ref{e2}) can be found in~\cite{qst1}.
As indicated by Eq. (\ref{e2}), a general unknown state $\rho_{in}$ could be reconstructed
by a series of projective measurements onto the eigenstates of $R(g)$.
When the tomographic group $G$ is selected to be the SU(2) group,
the tomography scheme is called {\it SU(2) tomography}
(see ~\cite{qst1} for the details).

Additionally, standard tomography can be implemented with measurements of the
basis operators of the space of density matrices. For example, an unknown qubit state can be expressed using the Pauli matrices
\begin{equation}\label{e3}
\rho_{in}=\frac{I}{2}+\sum_{i=x,y,z}\sigma_i\;\mathrm{tr}(\rho_{in}\sigma_i).
\end{equation}
By measuring the three Pauli matrices, we could reconstruct an unknown two-dimensional state.

\subsection{Direct state tomography}
The original direct state tomography (DST) is based on weak measurements.
Suppose a system in unknown state $\rho_{in}$ is weakly measured by a pointer in state $|\phi_0\rangle$.
The observables being measured on the system are chosen in the set $\{A_i=|a_i\rangle\langle a_i|\}_i$.
Therein, states $\{|a_i\rangle\}_i$ compose an orthogonal basis of the system's Hilbert space.
After that measurement we project the system onto another orthogonal basis $\{|\psi_f\rangle\}_f$ of the system's Hilbert space,
a process conventionally called {\it postselection} ( In this article, {\it postselection} represents the final projective measurement, and no data is discarded in the state reconstruction process.).
We then record the outcome of the postselection, read the pointer
by the pointer observable denoted as $\hat{s}$, and generate the
the {\it weak value} of each $A_i$ defined as \cite{dsm3}
\begin{equation}\label{e4}
W_{if}=\frac{\langle \psi_f|a_i\rangle\langle a_i|\rho_{in}|\psi_f\rangle}{P_f},
\end{equation}
where $P_f$ is the probability of obtaining $|\psi_f\rangle$ in postselection.
For convenience, the measurement basis $\{|a_i\rangle\}_i$
and the postselection basis $\{|\psi_f\rangle\}_f$ are chosen to be the mutually
unbiased bases (MUB) that $|\langle \psi_f|a_i\rangle|=1/\sqrt{d}$~\cite{mub}.
Using the weak values $\{W_{if}\}_{i,f}$,
we can reconstruct the unknown state using the formula
\begin{equation}
\rho_r = \sum_{i,f=1}^d \frac{P_f  W_{if}}{\langle \psi_f|a_i\rangle} |a_i\rangle\langle\psi_f|.
\label{e5}
\end{equation}
This formula states that no samples are
discarded, which is different with the original pure-state tomography case
in Ref. [11]. In the direct state tomography strategy of this article, each sample is used to construct weak values $\{W_{if}\}_{i,f}$ no matter which of $\{|\psi_f\rangle\}_f$ is obtained in postselection.

In this paper, we consider the setup of \cite{dsm1} where the pointer is a qubit
initialized in pure state $\rho_\phi=|0\rangle\langle 0|$,
the eigenstate of $\sigma_z$: $\sigma_z|0\rangle=|0\rangle$.
The weak measurement is described by the unitary coupling
\begin{equation}
U_i(g)=\exp(-ig A_i\otimes\sigma_x),
\label{uni}
\end{equation}
where the coupling strength $g$ is small.
The weak value is determined via
measuring two different observables on the pointer, i.e., $\hat{s}\in\{\hat{q},\hat{p}\}$,
$\hat{q}$ for the real part and $\hat{p}$ for the imaginary part.
They could be
\begin{equation}\label{s}
\hat{q}=\sigma_y, \qquad
\hat{p}=\sigma_x.
\end{equation}
The weak value is determined by
\begin{equation}
P_f W_{if}=\lim_{g\rightarrow 0}\frac{1}{2g} \mathrm{tr}\{
U_i(g)\rho_{in}\otimes\rho_\phi U_i^\dagger(g) \Pi_f\otimes(-\hat{q}+i\hat{p}) \}
\label{dsm},
\end{equation}
where $\Pi_f=|\psi_f\rangle\langle\psi_f|$. That is,
the weak value information is obtained by
measuring $\Pi_f\otimes\hat{s}$ on the joint system.

The validity of Eq. (\ref{dsm}) requires $g\rightarrow 0$,
which implies little information gain \cite{zhu} and low efficiency for DST.
However, $g$ is small but finite in an actual experiment,
thus a systematical error in the reconstruction is unavoidable.
These two disadvantages of DST have been verified by numerical simulations in~\cite{mac}.

\section{Modified direct state tomography}\label{sec3}
A recent work shows that weak value can be obtained exactly with
measurements of any strength, if we measure the {\it coupling-deformed} (CD) pointer
observables given in\cite{zhan}.
In this section, we will use this strategy to propose a modified direct state tomography (MDST),
and identify the optimal coupling strength which maximizes the efficiency of MDST.

\subsection{MDST with CD pointer observables}

When $g\to 0$, we measure the observable $\hat{s}\in{\hat{p},\hat{q}}$
given by Eq. (\ref{s}) on the pointer to
get weak value information.
In Ref. \cite{zhan}, we can obtain the
exact weak value information by measuring the CD
observable $\hat{s}(g)$ that varies with $g$, instead of $\hat{s}$.
For the two observables $\hat{p}$ and $\hat{q}$ in Eq. {\ref{s}},
using the method in~\cite{zhan},
the corresponding CD pointer observables $\hat{q}(g)$ and $\hat{p}(g)$ are calculated to be
\begin{equation}
\begin{aligned}
& \hat{q}(g)=\frac{1}{\sin g}\left(\sigma_y-\tan(\frac{g}{2})(I-\sigma_z)\right);\\
& \hat{p}(g)=\frac{1}{\sin g}\sigma_x.
\end{aligned}
\end{equation}
Then the weak value information is exactly obtained via
\begin{equation}\label{nwv}
P_f W_{if}=\frac{1}{2} \mathrm{tr}\{
U_i(g)\rho_{in}\otimes \rho_\phi U_i^\dagger(g)  \Pi_f\otimes[-\hat{q}(g)+i\hat{p}(g)]
\}.
\end{equation}
With the set $\{P_fW_{if}\}_{(i,f)}$, $\rho_{in}$ can be reconstructed via Eq. (\ref{e5}).
We call this scheme modified direct state tomography (MDST).

By comparing Eq. (\ref{dsm}) and Eq. (\ref{nwv}),
it is straightforward to see that in MDST, no approximation is applied, thus there will
be no inherent bias in the reconstruction, and MDST can be implemented with
any value of $g$.

Here, we would like to remark that our presentation of DST and MDST is slightly different
from others focusing on AAV's weak value, for example \cite{dsm7}.
If using the whole weak value, there will be an unknown normalization factor that can be fixed
only after all the measurements are completed \cite{vall}.
This is seen as conflicting with the claim of directness \cite{gro}.
However, in our MDST, Eq. (\ref{e5}) clearly shows that one element is directly determined
by $P_fW_{if}$,
which could be obtained from the measurements described in Eq. (\ref{nwv}).

\subsection{The optimal measurement strength in MDST}

Since in MDST $g$ can be any value, we need to search for
the optimal strength to obtain the highest efficiency.
First, we derive the optimal $g$ by statistical theory.
In actual implementation, MDST suffers from statistical errors.
This is the reason why there is a discrepancy
between the true state $\rho_t$ and the reconstructed state $\rho_r$.
Statistical errors can be quantified by the variance of the measured results.
Lower variance means less random error and higher reconstruction accuracy.
We find that when gauging the performance of MDST by the variance of the reconstruction,
the optimal value of $g$ will be appealingly state-independent.

\begin{figure}[t]
    \centering \includegraphics[scale=0.4]{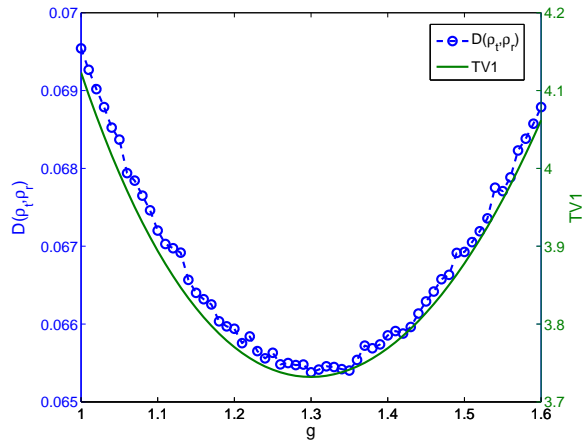}
    \caption{(Color online) The values of TV1 (green line) of the reconstruction,
    and trace distance $D(\rho_t, \rho_r)$ (blue circles) averaged over $10^5$ states,
    against the coupling strength.}  \label{f1}
\end{figure}

From Eq. (\ref{e5}), each element of the reconstruction $\rho_r$
\begin{equation}\label{reconstruction}
\varrho_{if}={P_f  W_{if}}/{\langle \psi_f|a_i\rangle}
\end{equation}
is determined by two independent measurements, one for the real part and
the other for the imaginary part of $\varrho_{if}$.
Therefore, the total variance of the reconstruction state $\rho_r$ can be defined as
\begin{equation}
\delta^2 \rho_r= \sum_{i,f=1}^d \delta^2(\Re\varrho_{if})+\delta^2(\Im\varrho_{if}),
\end{equation}
where $\Re$ and $\Im$ stand for the real and imaginary part, respectively.
From Eqs. (\ref{e5}) and (\ref{nwv}), the convention $|\langle \psi_f|a_i\rangle|=1/\sqrt{d}$,
and the error propagation theory \cite{error},
the total variance $\delta^2 \rho_r$ can be expressed as
\begin{equation}
\begin{aligned}
\delta^2\rho_r=\frac{d}{4} & \sum_{\hat{s}\in\{\hat{q},\hat{p}\}} \sum_{i,f=1}^d
 \mathrm{tr} \{ U_i \rho_t\otimes\rho_\phi U_i^\dagger \Pi_f\otimes\hat{s}^2(g)\}\\
& -\left(\mathrm{tr}\{ U_i \rho_t\otimes\rho_\phi U_i^\dagger \Pi_f\otimes\hat{s}(g)\}\right)^2,
\end{aligned}
\label{var}
\end{equation}
As shown in Eq. (\ref{nwv}), the second term in the summation gives the squares of the real
or imaginary parts of $\varrho_{if}$.
Since it is independent of $g$,
we only need to focus on the summation of the first term of Eq. (\ref{var}),
which will be denoted as $TV1$.
Using the relation that $\sum_f \Pi_f=I_s$ (the identity on the system's Hilbert space),
we have
\begin{equation}
TV1=\frac{d}{4} \sum_{\hat{s}\in\{\hat{p},\hat{q}\}}\sum_{i=1}^d \mathrm{tr} \{\rho_i\hat{s}^2(g)\}
\label{sumf}
\end{equation}
where $\rho_i=\mathrm{tr}_s(U_i \rho_t\otimes\rho_\phi U_i^\dagger)$.
Since $\hat{p}(g)^2=1/\sin^2(g)$, we obtain
\begin{equation}
\mathrm{tr} \{ \rho_i \hat{p}^2(g)\}=\frac{1}{\sin^2(g)},
\end{equation}
which is independent on the input state $\rho_t$.
For $\hat{s}(g)=\hat{q}(g)$, we have
\begin{equation}
\begin{aligned}
\mathrm{tr}_a & \{\rho_i\hat{q}^2(g)\} = \frac{1}{\sin^2 g}+\frac{2c_i}{\cos^2\frac{g}{2}},
\end{aligned}
\end{equation}
where $c_i=\langle a_i|\rho_t|a_i\rangle$.
Since $\sum_i c_i=1$, TV1 is evaluated as
\begin{equation}\label{e17}
TV1=\frac{d^2}{2\sin^2 g}+\frac{d}{2\cos^2\frac{g}{2}},
\end{equation}
where $d$ is the dimension of the unknown state.
TV1, and thus $\delta^2\rho_r$, attains the minimum at the optimal
strength
\begin{equation}\label{e18}
g_{opt}=\arccos(1+\frac{d}{2}-\sqrt{d+\frac{d^2}{4}}).
\end{equation}
In Fig. \ref{f1}, the value of TV1 against the coupling strength $g\in[1,1.6]$
is illustrated for the two-dimensional systems.
From Eq. (\ref{e18}), the minimum of $TV1$ locates at $g=1.30$.
It may be surprising at first glance that $g=\pi/2$, which makes
the coupling unitary Eq. (\ref{uni}) describe the strongest measurement,
is not optimal.
Our result $g_{opt}=1.3$ is also consistent with the result of
$g_{opt}\approx 1.25$ \cite{gro}, which is calculated using the average fidelity
as the figure of merit of the reconstruction of the pure qubit states in the scheme introduced by Vallone and Dequal~\cite{vall}.

In order to align with the results in \cite{mac},
we will use the trace distance to quantify the accuracy of the
reconstruction.
To locate the optimal strength with respect to trace distance,
we use the standard Monte Carlo method to simulate MDST:
first, a two-dimensional mixed state $\rho_t$ is selected randomly
as the target state; second, the reconstruction $\rho_r$ is produced
by MDST with $10^3$ copies of $\rho_t$ for different coupling strengths;
third, the trace distance~\cite{niel}
between $\rho_t$ and  $\rho_r$,
\begin{equation}\label{e25}
 D(\rho_t,\rho_r)=\frac{\mathrm{tr}\left(|\rho_t-\rho_r|\right)}{2},
\end{equation}
is calculated to gauge the performance of MDST. Smaller trace
distance $D(\rho_t, \rho_r)$ means higher efficiency.
In order to eliminate statistical fluctuations, we have averaged
$D(\rho_t, \rho_r)$ over $10^5$ randomly selected $\rho_t$.
The simulation results are presented in Fig. \ref{f1}.
It coincides well with the prediction of variance that $g_{opt}=1.3$.

\section{Comparison}\label{sec4}
The value of TV1 diverges at the weak limit $g\rightarrow 0$.
This demonstrates the conclusion that DST at the weak limit suffers serious
random noise, which leads to low efficiency.
The exactness of the MDST formalism, and its validity over the entire
range of measurement strength suggest that MDST can overcome the problems
of low efficiency and intrinsic bias.
In this section, we will examine this claim by using Monte Carlo simulations
to compute the three strategies of quantum state reconstruction:
MDST, DST, and SU(2) tomography.

The simulation goes as follows.
First, a state is reconstructed by MDST, DST, Pauli tomography and SU(2) tomography;
second, the trace distance $D(\rho_t,\rho_r)$ between the true state $\rho_t$ and the
reconstructed state $\rho_r$ is used to gauge the estimation efficiency.
Given equivalent sample size,
the smaller the trace distance is,
the higher the efficiency of a tomography scheme will be.

\begin{figure}[t]
\centering \includegraphics[scale=0.4]{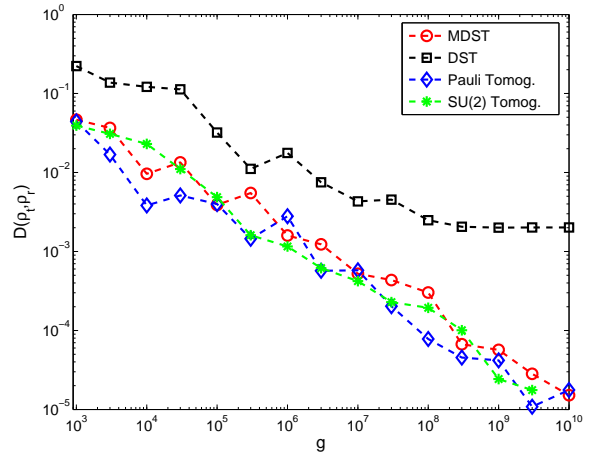} \caption{(Color online) Comparison of the efficiencies of MDST, DST, SU(2) tomography and Pauli tomography for qubits. The trace distance $D(\rho_t,\rho_r)$ is plotted as the function of the number of copies N of the system. Circles, D calculated from MDST with $g=1.3$; squares, D calculated from DST with $g=0.1$; diamonds, D calculated from Pauli tomography; stars, D calculated from SU(2) tomography.}  \label{f2}
\end{figure}

As shown in Fig. \ref{f2}, all the reconstruction strategies are affected by statistical errors.
The trace distances decrease with the the numbers of the copies of the systems.
As expected from the simulation results, the reconstruction $\rho_r$ of DST has a systematical bias, while the state $\rho_r$ obtained by MDST has no error bias. The trace distance $D(\rho_t,\rho_r)$ continues to decrease as the number of copies increases.

\begin{figure}[t]
    \centering \includegraphics[scale=0.4]{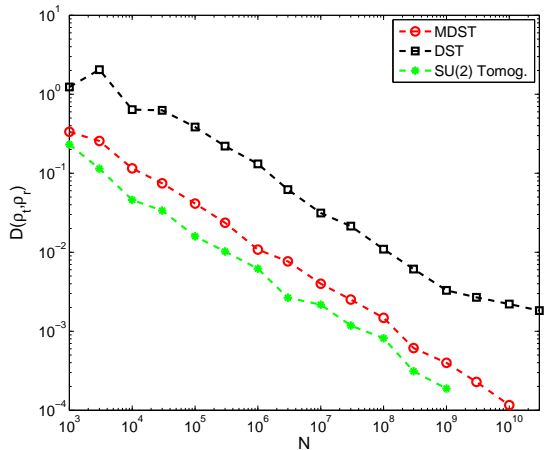}
    \caption{(Color online) Comparison of MDST, DST and SU(2) tomography for five-dimensional states. The trace distance $D(\rho_t,\rho_r)$ is plotted as the function of the number of copies N of the system. Circles, D calculated from MDST with $g=1.4$; squares, D calculated from DST with $g=0.1$; stars, D calculated from SU(2) tomography.}  \label{f3}
\end{figure}

For quibts, as indicated in Fig. \ref{f2}, the efficiencies of MDST, Pauli tomography and SU(2) tomography differ little.
For five-dimensional systems, as shown in Fig. \ref{f3}, the relationship between the trace distance and the number of copies is given. These two figures clearly show that to reach the same level of trace distance,
MDST uses far fewer samples than DST.
That is to say, the efficiency of DST is significantly improved by using stronger measurements and CD pointer observables.

Figure \ref{f3} also indicates that MDST is less efficient than SU(2) tomography.
For the same reconstruction precision, MDST needs more copies than SU(2) tomography.
This is consistent with the result found in Ref. \cite{gro} that the scheme of Haar-uniform randomly
chosen one-dimensional orthogonal projective measurements is more efficient
than direct tomography methods.

In order to study the gap between the efficiencies of MDST and SU(2) tomography,
we have performed further simulations for two, four, five, six, eight, nine, and ten-dimensional systems.
In the tables of the appendix, we list the reconstruction precision gauged by
trace distance, and the corresponding samples size
$N_{MDST}$ and $N_{SU(2)}$, which are averaged over 500 repeated reconstructions to decrease the statistical fluctuation.

As indicated in Tables I-III, for the expected trace distances $D$,
the results suggest that $N_{MDST} \approx 0.8 d N_{SU(2)}$ for two,
four and five-dimensional systems, where $d$ is the dimension of the systems.
In Tables IV-V, for six and eight-dimensional states, it is shown that $N_{MDST} \approx 0.75 d N_{SU(2)}$.
In Tables VI-VII, $0.7 d N_{SU(2)}$ copies are needed in MDST to attain the same
precision of SU(2) tomography for nine and ten-dimensional systems.
Roughly speaking, we suppose $N_{MDST}\approx 0.8 d N_{SU(2)}$.

From the data obtained in simulations,
we estimate that to reach an equivalent level of reconstruction accuracy,
the sample size required in MDST, $N_{MDST}$,
is about $0.8 d$ times of $ N_{SU(2)}$, the sample size required in SU(2) tomography.
MDST is clearly less efficient than SU(2) tomography, especially for high dimensional systems.

Since SU(2) tomography is predicted on measuring a complete set of non-commuting observables,
a difficult task to realize in actual experiments.
Although MDST is less efficient than SU(2) tomography,
MDST is much easier to implement in experiments.
MDST might be more useful than SU(2) tomography for reconstructing an unknown state,
especially for high dimensional states.

\section{conclusion}\label{sec5}

In this paper, we have presented a modified direct state tomography (MDST)
using the coupling-deformed pointer observables.
We have verified that MDST has no inherent bias.
MDST is valid for any large coupling strength.
We have obtained the optimal measurement strength
with which the efficiency of MDST is much higher than that of DST.
Numerical simulation also suggests that the efficiency of MDST
is less than SU(2) tomography.
However, MDST is much easier to implement in actual experiments, and it thus could be useful in reconstructing unknown quantum states.

\section*{Acknowledgments}
We thank Lorenzo Maccone for his helpful suggestions on SU(2) tomography. This work was financially supported by the National Natural Science Foundation of China (Grants No. 11305118, and No. 11475084), and the Fundamental Research Funds for the Central Universities.

\section*{Appendix}
In this appendix, we list seven tables to present the state reconstruction precision levels and the numbers  needed in MDST and SU(2) tomography respectively. All the trace distances are averaged over $10^3$ repeated reconstructions to eliminate statistical fluctuations.

\begin{table}[!htbp]
\begin{tabular}{|c|c|c|c|}
\hline
\multirow{2}{*}{\shortstack{MDST \\ $g=1.3$} } & $N_1=1600$ & $N_2=2560$ & $N_3=4096$ \\
\cline {2-4} & $D_1=0.0497$ & $D_2=0.0386$ & $D_3=0.0300$ \\
\hline
\multirow{2}{*}{\shortstack{SU(2)\\tomography}} & $N_1'=1000$ &$N_2'=1600$ & $N_3'=2560$ \\
\cline {2-4} & $D_1'=0.0499$ & $D_2'=0.0384$ & $D_3'=0.0299$ \\
\hline
\end{tabular}

\caption{The trace distances $D$ ($D'$) and the number $N$ ($N'$) of copies needed for two-dimensional systems in MDST (SU(2) tomography). }
\end{table}

\begin{table}[!htbp]
\begin{tabular}{|c|c|c|c|}
\hline
\multirow{2}{*}{\shortstack{MDST \\ $g=1.4$} } & $N_1=3200$ & $N_2=10240$ & $N_3=32768$ \\
\cline {2-4} & $D_1=0.140$ & $D_2=0.0780$ & $D_3=0.0435$ \\
\hline
\multirow{2}{*}{\shortstack{SU(2)\\tomography}} & $N_1'=1000$ &$N_2'=3200$ & $N_3'=10240$ \\
\cline {2-4} & $D_1'=0.141$ & $D_2'=0.0784$ & $D_3'=0.0435$ \\
\hline
\end{tabular}

\caption{The trace distances $D$ ($D'$) and the number $N$ ($N'$) of copies needed for four-dimensional systems in MDST (SU(2) tomography).}
\end{table}

\begin{table}[!htbp]
\begin{tabular}{|c|c|c|c|}
\hline
\multirow{2}{*}{\shortstack{MDST \\ $g=1.4$} } & $N_1=4000$ & $N_2=16000$ & $N_3=64000$ \\
\cline {2-4} & $D_1=0.192$ & $D_2=0.0961$ & $D_3=0.0481$ \\
\hline
\multirow{2}{*}{\shortstack{SU(2)\\tomography}} & $N_1'=1000$ &$N_2'=4000$ & $N_3'=16000$ \\
\cline {2-4} & $D_1'=0.196$ & $D_2'=0.0977$ & $D_3'=0.0482$ \\
\hline
\end{tabular}
\caption{The trace distances $D$ ($D'$) and the number $N$ ($N'$) of copies needed for five-dimensional systems in MDST (SU(2) tomography).}
\end{table}

\begin{table}[!htbp]
\begin{tabular}{|c|c|c|c|}
\hline
\multirow{2}{*}{\shortstack{MDST \\ $g=1.4$} } & $N_1=9000$ & $N_2=40500$ & $N_3=182250$ \\
\cline {2-4} & $D_1=0.182$ & $D_2=0.0849$ & $D_3=0.0404$ \\
\hline
\multirow{2}{*}{\shortstack{SU(2)\\tomography}} & $N_1'=2000$ &$N_2'=9000$ & $N_3'=40500$ \\
\cline {2-4} & $D_1'=0.184$ & $D_2'=0.0864$ & $D_3'=0.0409$ \\
\hline
\end{tabular}

\caption{The trace distances $D$ ($D'$) and the number $N$ ($N'$) of copies needed for six-dimensional systems in MDST (SU(2) tomography).}
\end{table}

\begin{table}[!htbp]
\begin{tabular}{|c|c|c|c|}
\hline
\multirow{2}{*}{\shortstack{MDST \\ $g=1.4$} } & $N_1=18000$ & $N_2=108000$ & $N_3=648000$ \\
\cline {2-4} & $D_1=0.224$ & $D_2=0.0916$ & $D_3=0.0374$ \\
\hline
\multirow{2}{*}{\shortstack{SU(2)\\tomography}} & $N_1'=3000$ &$N_2'=18000$ & $N_3'=108000$ \\
\cline {2-4} & $D_1'=0.233$ & $D_2'=0.0954$ & $D_3'=0.0390$ \\
\hline
\end{tabular}

\caption{The trace distances $D$ ($D'$) and the number $N$ ($N'$) of copies needed for eight-dimensional systems in MDST (SU(2) tomography). }
\end{table}

\begin{table}[!htbp]
\begin{tabular}{|c|c|c|c|}
\hline
\multirow{2}{*}{\shortstack{MDST \\ $g=1.4$} } & $N_1=31500$ & $N_2=94500$ & $N_3=315000$ \\
\cline {2-4} & $D_1=0.211$ & $D_2=0.123$ & $D_3=0.0677$ \\
\hline
\multirow{2}{*}{\shortstack{SU(2)\\tomography}} & $N_1'=5000$ &$N_2'=15000$ & $N_3'=50000$ \\
\cline {2-4} & $D_1'=0.214$ & $D_2'=0.126$ & $D_3'=0.0683$ \\
\hline
\end{tabular}

\caption{The trace distances $D$ ($D'$) and the number $N$ ($N'$) of copies needed for nine-dimensional systems in MDST (SU(2) tomography).}
\end{table}

\begin{table}[!htbp]
\begin{tabular}{|c|c|c|c|}
\hline
\multirow{2}{*}{\shortstack{MDST \\ $g=1.4$} } & $N_1=21000$ & $N_2=70000$ & $N_3=210000$ \\
\cline {2-4} & $D_1=0.316$ & $D_2=0175$ & $D_3=0.100$ \\
\hline
\multirow{2}{*}{\shortstack{SU(2)\\tomography}} & $N_1'=3000$ &$N_2'=10000$ & $N_3'=30000$ \\
\cline {2-4} & $D_1'=0.327$ & $D_2'=0.180$ & $D_3'=0.103$ \\
\hline
\end{tabular}

\caption{The trace distances $D$ ($D'$) and the number $N$ ($N'$) of copies needed for ten-dimensional systems in MDST (SU(2) tomography). }
\end{table}

\end{document}